\documentclass[twocolumn,amsmath,amssymb]{snp}
\usepackage{graphicx}
\usepackage{dcolumn}
\usepackage{bm}
\newcommand{\nn}{\nonumber}
\newcommand{\be}{\begin{equation}}
\newcommand{\ee}{\end{equation}}
\newcommand{\bea}{\begin{eqnarray}}
\newcommand{\eea}{\end{eqnarray}}

\begin{document}

\title{{\Large  Mapping QGP interaction through its temperature dependent degeneracy factor}}

\author{\large Ranjesh Kumar$^{1,*}$, Ankit Anand$^{1}$, Souvik Paul$^{1}$, Sarthak Satapathy$^2$,
Sabyasachi Ghosh$^2$}
\email{rk17ms029@iiserkol.ac.in}
\affiliation{$^1$Department of Physical Sciences,Indian Institute of Science Education 
and Research Kolkata, Mohanpur, West Bengal 741246, India}
\affiliation{$^3$Indian Institute of Technology Bhilai, GEC Campus, Sejbahar, Raipur 492015, 
Chhattisgarh, India}
\maketitle

Present work has attempted to map interaction picture of quark gluon plasma (QGP)
in terms of its temperature dependent degeneracy factor. 
From standard statistical mechanical description, entropy density $s$ of massless QGP can be obtained as
\be
s=\Big[g_g+(g_u+g_s)\Big(\frac{7}{8}\Big)\Big]\frac{4\pi^2}{90}T^3\approx 20.8~T^3~,
\label{s_QGP_m0}
\ee
which is popularly known as Stephan-Boltzmann (SB) limits.
\begin{figure}
	\centering
	\includegraphics[scale=0.27]{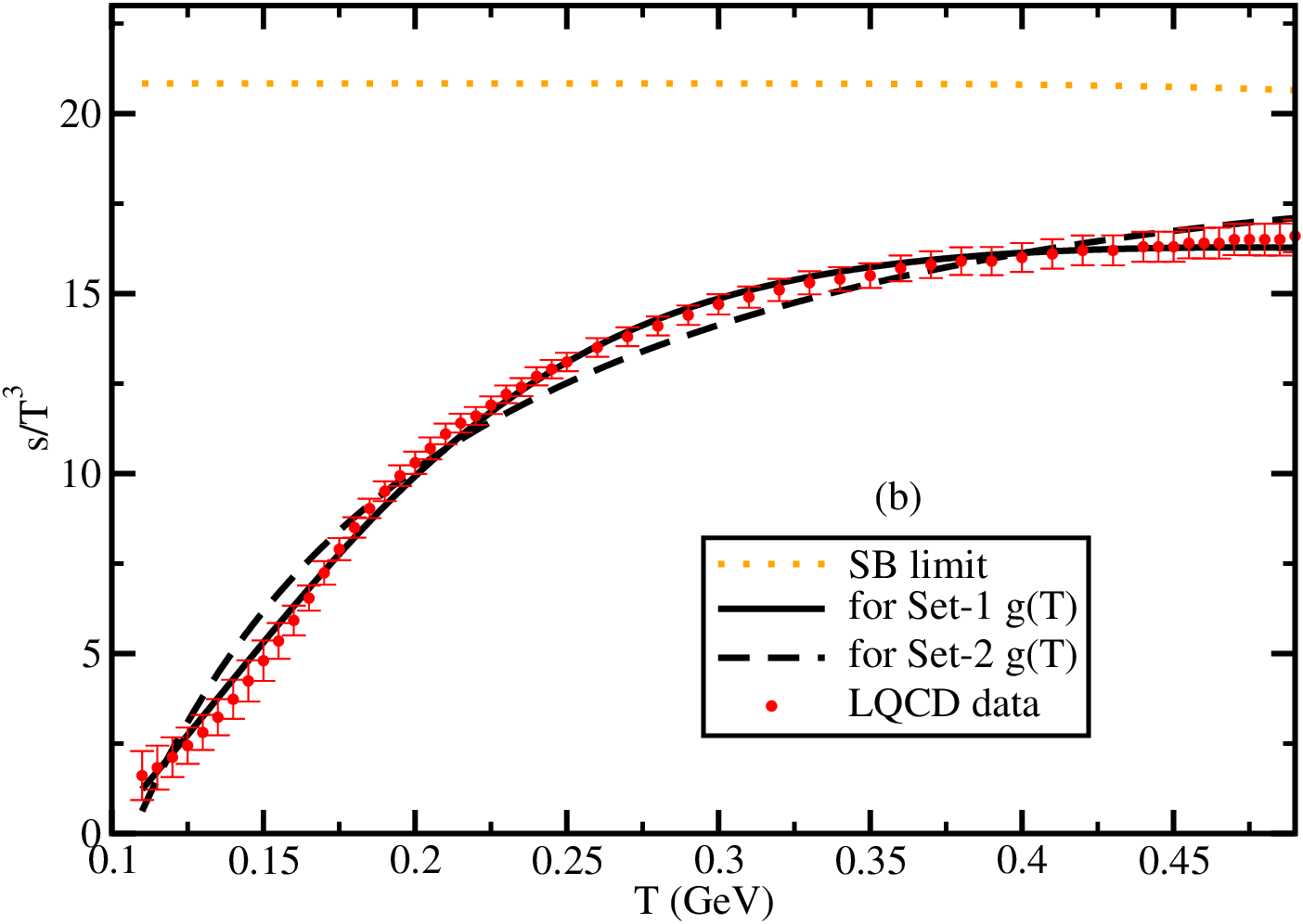} 
	\caption{LQCD data of $s/T^3$ from Refs.~\cite{LQCD1,LQCD2} shown by red circles
	with error bars, straight horizontal (brown) dotted line indicates SB limits of $s/T^3$.
	LQCD data has been matched by using two sets of temperature dependent 
	degeneracy factor $g(T)$.}
	\label{svsT}
\end{figure}
According to lattice Quantum Chromo Dynamics (LQCD) calculation~\cite{LQCD1,LQCD2}, 
the numerical values of $s$ for QGP remain
always lower than its SB limits. In Fig.~(\ref{svsT}), red circles present the LQCD
data of $s/T^3$ from Refs.~\cite{LQCD1,LQCD2} and brown dotted line indicates its SB limit. 
It is the interaction of the system, for which
$s$ is reduced from its SB or non-interaction limit.
By assuming appropriate temperature dependent degeneracy
factors of quarks and gluons, one can construct LQCD data points for $s(T)$.
For this purpose, we have considered a temperature dependent factor $g(T)$, 
attached with $g_{q,s,g}$ in Eq.~(\ref{s_QGP_m0}) and then match the LQCD 
data of $s(T)$~\cite{LQCD1,LQCD2}.
We get a parametrized expression:
\be
g(T)= a_0 - \frac{a_1}{e^{a_2(T-a_3)}+ a_4}~,
\label{gT_1}
\ee
where $a_0 = 0.793$, $a_1 = 0.687$, $a_2 = 16.284$, $a_3=0.170$, $a_4 = 0.560$~.
\begin{figure}
	\centering
	\includegraphics[scale=0.27]{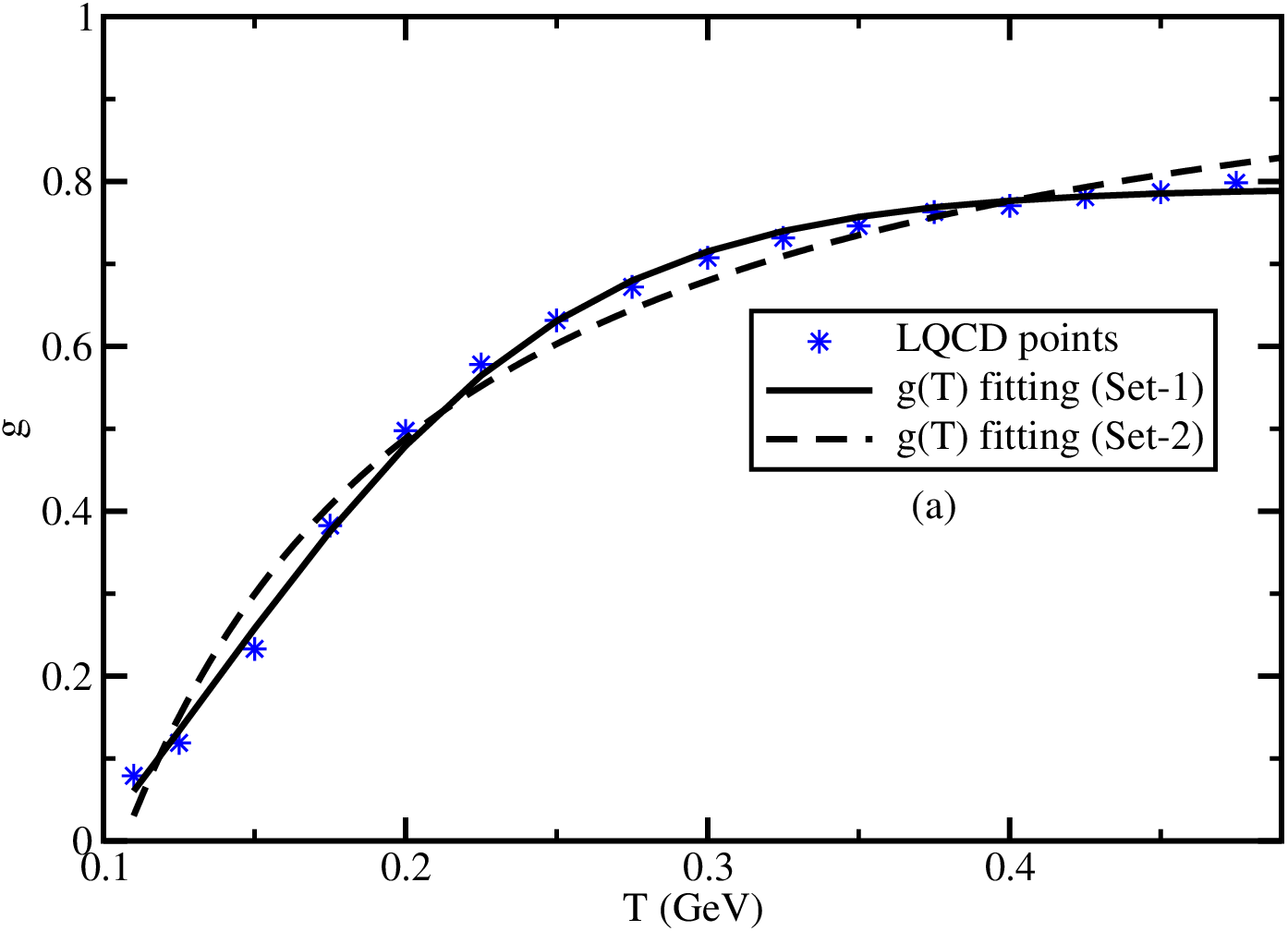} 
	\caption{Temperature dependence degeneracy factors $g(T)$ parametrization curves - Set-1 (solid line),
	Set-2 (dash line) and LQCD extracted points (stars).}
	\label{gvsT}
\end{figure}
The above set of parameters (say set-1) provide better matching to LQCD data but it is not satisfying the expectation
of reaching SB limit of $s$ at $T\rightarrow\infty$.
To fulfill the condition, we have restricted $a_0=1$, and get the another parametrized function:
\be
g(T)= 1 - \frac{b_0}{e^{b_1(T-b_2)}+ b_3}~,
\label{gT_2}
\ee
where $b_0 = 0.793$, $b_1 = 0.687$, $b_2=0.170$, $b_3 = 16.284$, which can be called set-2.
%

Fig.~\ref{gvsT} shows two set of $g(T)$ (dash and solid lines) and LQCD data points (stars)~\cite{LQCD1,LQCD2}. 
Their corresponding
values of $s/T^3$ is plotted in Fig.~(\ref{svsT}).
So effectively total degeneracy factor of QGP $g_u+g_s+g_g=52$ will be suppressed for considering $g(T)*g_{u,s,g}$,
and around $T=0.200$ GeV, $g(T)\approx 0.5$, which means effective degeneracy factor become $0.5\times 52=26$.
In hadronic temperature range, around $T\approx 0.120$ GeV, $g(T)\approx 0.13$ will provide effective degeneracy
factor $0.13\times 52=7$, which is exactly hadronic degeneracy factor $g_\pi +g_K=7$. 
\begin{figure}
	\centering
	\includegraphics[scale=0.27]{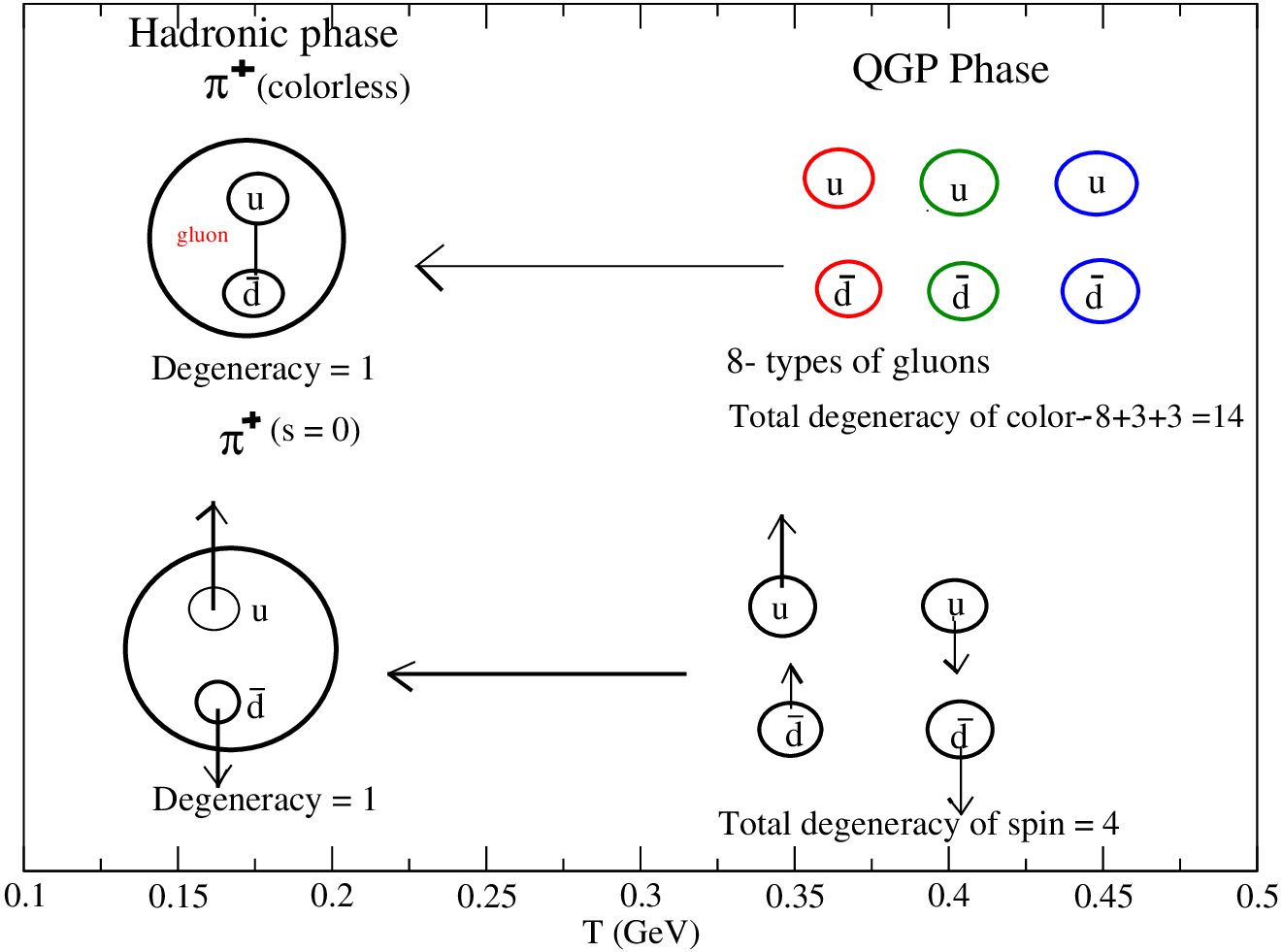} 
	\caption{A schematic diagram shows two phases - QGP and hadronic matter, which
	carry different number of color (upper part) and spin (lower part) degeneracy factors.}
	\label{Deg_QGP}
\end{figure}
So in this way, we might
roughly map QCD interaction picture via shrinking of degeneracy factor of quarks and gluons with lowering 
the temperature. This fact can be compared with the fact of temperature dependent 
degree of freedom for di-atomic or n-atomic molecule. At low temperature degrees of freedoms
of di-atomic or n-atomic molecules is $3\times 2-1$ or $3\times n - k$ because of its 1 or k
number of atomic bondings, which can be broken at high temperature and degrees of freedom enhanced
as
\bea
3\times 2-1=5 &\rightarrow& 3\times 2=6
\nn\\
{\rm or},
\nn\\
3\times n - k &\rightarrow& 3\times n~.
\eea
According to equipartition theorem of thermodynamics, internal energy of 
di-atomic or n-atomic molecular system will be proportional to its degrees 
of freedom, hence internal energy (other thermodynamical quantities) will also
be increased with increasing temperature.
To visualize the similar kind of temperature dependence of degeneracy factor
for QGP system, we have drawn a schematic diagram in Fig.~(\ref{Deg_QGP}). 
We have sketched two different number of color degeneracy factors for high $T$ QGP phase 
and low $T$ hadrinic phase in the upper part of Fig.~(\ref{Deg_QGP}), while same for 
spin degeneracy factors is drawn in lower part of Fig.~(\ref{Deg_QGP}). In QGP phase,
one can identify 8 color gluons, 3 color $u$ and 3 color ${\bar d}$, which can form
a color-less $\pi^+$ state in low $T$ hadronic phase. This reduction
of $8+3+3=14$ color states to one color-less states may be considered one of the possibility
however, collective reduction from $g=52$ at $T\rightarrow\infty$ to $g\approx 7$
at $T=0.120$ GeV is occurring through interaction. Hence, through temperature dependent
degeneracy factor, this interaction is attempted to map in present work~\cite{LQCD_QGP}.


\begin{thebibliography}{50}
%
\bibitem{LQCD1} S. Borsanyi et. al.,
{\it Full result for the QCD equation of state with 2+1 flavors},
Phys. Lett. {\bf B 370} (2014) 99, arXiv:1309.5258 [hep-lat].
%
\bibitem{LQCD2}A. Bazavov et al.
{\it Equation of state in (2+1)-flavor QCD},
Phys. Rev. D 90, 094503 (2014)
%
\bibitem{LQCD_QGP} S. Satapathy, S. Paul, A. Anand, R. Kumar, S. Ghosh,
{\it From Non-interacting to Interacting Picture of Thermodynamics 
and Transport Coefficients for Quark Gluon Plasma},
arXiv:1908.04330 [hep-ph]
%
\end{thebibliography}
\end{document}